\documentclass[a4paper,11pt]{article}
\pdfoutput=1 % if your are submitting a pdflatex (i.e. if you have
\usepackage{jcappub} % for details on the use of the package, please
\usepackage[T1]{fontenc} % if needed

\def\beq{\begin{equation}}
\def\eeq{\end{equation}}
\def\bma{\begin{math}}
\def\ema{\; \end{math}}
\def\begary{\beq\begin{array}{c}}
\def\endary{\end{array}\eeq}
\def\begaryl{\beq\begin{array}{l}}
\def\endaryl{\end{array}\eeq}

\def\bmat{\left( \begin{array}{ccc}}
\def\emat{\end{array}\right)}
\def\bmatv{\left( \begin{array}{cccc}}
\def\ematv{\end{array}\right)}
\def\bmatz{\left( \begin{array}{cc}}
\def\ematz{\end{array}\right)}
\def\bvec{\left( \begin{array}{c}}
\def\evec{\end{array}\right)}
\def\percent{\%}
\def\X2{$\chi^2$}

\title{\boldmath CALET's Sensitivity to Dark Matter Annihilation in the Galactic Halo}

\author[a,1]{H. Motz,\note{Corresponding author.}}
\author[b]{Y. Asaoka,}
\author[b]{S. Torii,}
\author[c]{S. Bhattacharyya}

\affiliation[a]{International Center for Science and Engineering Projects, Waseda University,\\3-4-1 Okubo, Shinjuku, Tokyo 169-8555, Japan}
\affiliation[b]{Research Institute for Science and Engineering, Waseda University,\\3-4-1 Okubo, Shinjuku, Tokyo, 169-8555, Japan}
\affiliation[c]{Advanced School for Science and Engineering, Waseda University,\\3-4-1 Okubo, Shinjuku, Tokyo, 169-8555, Japan}

\emailAdd{motz@aoni.waseda.jp}
\emailAdd{yoichi.asaoka@aoni.waseda.jp}
\emailAdd{torii.shoji@waseda.jp}
\emailAdd{saptashwab@ruri.waseda.jp}

\abstract{CALET (Calorimetric Electron Telescope), installed on the ISS in August 2015, directly measures the electron+positron cosmic rays flux up to 20 TeV. 
With its proton rejection capability of $1 : 10^5$ and an aperture of 1200 cm$^2\cdot$sr, it will provide good statistics even well above one TeV, while also featuring an energy resolution of 2\%, which allows it to detect fine structures in the spectrum. Such structures may originate from Dark Matter annihilation or decay, making indirect Dark Matter search one of CALET's main science objectives among others such as identification of signatures from nearby supernova remnants, study of the heavy nuclei spectra and gamma astronomy. 
The latest results from AMS-02 on positron fraction and total electron+positron flux can be fitted with a parametrization including a single pulsar as an extra power law source with exponential cut-off, which emits an equal amount of electrons and positrons. This single pulsar scenario for the positron excess is extrapolated into the TeV region and the expected CALET data for this case are simulated. Based on this prediction for CALET data, the sensitivity of CALET to Dark Matter annihilation in the galactic halo has been calculated.
It is shown that CALET could significantly improve the limits compared to current data, especially for those Dark Matter candidates that feature a large fraction of annihilation directly into $e^+ + e^-$, such as the LKP (Lightest Kaluza-Klein particle). 

}

\begin{document}
\maketitle
\flushbottom
\section{Introduction}

%CALET also doubles as a gamma observatory with an energy resolution nearly as good as for charged cosmic rays, making it one of the best suited experiments to look for a line in the gamma spectrum from direct annihilation of Dark Matter into photons \cite{Yoshida-ICRC}.   

The CALET cosmic ray experiment  \cite{Torii-ICRC} can play an important role in indirect Dark Matter search, as it will provide the first direct measurement of the TeV region of electron+positron cosmic rays. With its high energy resolution, it has the best potential to detect structures originating from Dark Matter annihilation or decay, discerning them from astrophysical sources. Dark Matter search in the high-energy electron spectrum is distinguished from indirect Dark Matter search in the $\rm{\gamma}$-spectrum by the rather short propagation distance of electrons in the galaxy, which makes the expected signal rely on the local and nearby Dark Matter density, instead of the density at the Galactic Center or other distant overdense regions, such as dwarf galaxies. Measurement of the local Dark Matter density by star movement \cite{0004-637X-772-2-108} allows to determine the Dark Matter density within 1-2 kpc distance from Sun, matching the approximate range of electron propagation. The precision of this measurement improves steadily by observation and new analysis methods \cite{2015arXiv150708581S}, and gives an estimate of the density independent of the large uncertainty from the simulation-derived halo models. Direct search, while also tied to the local density, depends on the scattering of Dark Matter with nuclei and thus probes a different set of Dark Matter properties.  
On the other hand, looking for Dark Matter signatures in electron and positron cosmic rays requires to take into account the excess in the positron fraction, discovered by the PAMELA experiment \cite{Adriani:2008zr} and currently measured with highest statistics by AMS-02 \cite{PhysRevLett.113.121101}. The AMS-02 collaboration proposed the excess to be caused by a single power law extra source with an exponential cut-off \cite{PhysRevLett.110.141102}, emitting an equal flux of electrons and positrons. To investigate the sensitivity of CALET to Dark Matter annihilation signatures, the parametrization introduced by this approach is extended to the total electron+positron flux and to include propagation effects at higher energies.
The parametrization together with the AMS-02 total electron+positron flux up to one TeV \cite{PhysRevLett.113.221102} is used to extrapolate this scenario into the yet unknown TeV region, assuming that the extra component is purely from a nearby pulsar. The parametrization and choice of parameters are compared to numerical propagation calculations, showing that the chosen scenario is in agreement with generally accepted parameters for propagation and the injection spectra of the background flux and the extra source.
Based on this prediction, the expected CALET total flux measurement results for 5~years of data-taking were simulated. Adding a hypothetical additional component from Dark Matter, the limits that could be set on annihilation of several Dark Matter candidates have been calculated, under the condition that CALET data will match the pure pulsar case.

\section{Parametrization of the Local Cosmic Ray Spectrum}
\label{DMmeth}
To determine the minimum contribution from Dark Matter annihilation to the electron and positron flux which is undiscoverable by a given cosmic ray experiment against a variable background, a parametrization of the local flux with terms for all relevant cosmic ray components is required. It has to be sufficently precise to match current measurements to provide a valid background case, and should be compatible with generally accepted cosmic ray acceleration and propagation models.
To describe the local cosmic ray electron and positron spectrum with a primary component originating from distant supernovae and a secondary component from nuclei cosmic rays interacting with the interstellar medium in a parametrization, two power law indices $\gamma_{e}$ and $\gamma_{e^{+}}$ and two coefficients $C_{e}$ and $C_{e^{+}}$ for the combined electron+positron and positron-only flux are required. Radiative energy losses of this diffuse background spectrum are modeled as an exponential cut-off at around energy $E_{cut_d}$.

The scenario of a single pulsar as extra source is parametrized by a power law term for both electrons and positrons with index $\gamma_{s}$, coefficient $C_{s}$ and cut-off energy $E_{cut_s}$. For the purpose of studying CALET's sensitivity to it, a component from Dark Matter annihilation $\Phi_{DM}$ scaled by a Boost Factor $BF$ is introduced, so that the total flux is given by:
\small \beq \Phi_{e}(E) =  2 \Phi_{DM}(E) \cdot BF + C_{e} E^{\gamma_{e}} \left(2 {C_{s} \over C_{e}} E^{\gamma_{s}-\gamma_{e}} \cdot \exp\left({-E \over E_{cut_s}}\right) + \left({C_{e^{+}} \over C_{e}} \cdot E^{\gamma_{e^{+}}-\gamma_{e}}+1\right)  \cdot \exp\left({-E \over E_{cut_d}}\right)\right) \label{totflx}\eeq \normalsize

With both Dark Matter annihilation and the pulsar emitting an equal ratio of electrons and positrons, the positron fraction is calculated as: 

\beq {{\Phi_{e^{+}}(E) \over \Phi_{e}(E)} =}{ { {  {{\Phi_{DM}(E) \cdot BF} \over {C_{e} E^{\gamma_{e}} }} + {C_{s} \over C_{e}} E^{\gamma_{s}-\gamma_{e}} \cdot \exp\left({-E \over E_{cut_s}}\right) + {C_{e^{+}} \over C_{e}} \cdot E^{\gamma_{e^{+}}-\gamma_{e}} \cdot \exp\left({-E \over E_{cut_d}}\right)}         }     \over  {   {  {{2 \cdot {\Phi_{DM}(E) \cdot BF} \over {C_{e} E^{\gamma_{e}} }}} + 2{{C_{s}} \over C_{e}} E^{\gamma_{s}-\gamma_{e}} \cdot \exp\left({-E \over E_{cut_s}}\right) + \left({C_{e^{+}} \over C_{e}} \cdot E^{\gamma_{e^{+}}-\gamma_{e}}+1\right) \cdot \exp\left({-E \over E_{cut_d}}\right)} \label{posfract}       } } \eeq

% the positron flux is thus \beq \Phi_{e^{+}}(E) = \Phi_{DM}(E) \cdot BF + C_{e} E^{\gamma_{e}} \left({C_{s} \over C_{e}} E^{\gamma_{s}-\gamma_{e}} \cdot \exp\left({-E \over E_{cut_s}}\right) + {C_{e^{+}} \over C_{e}} \cdot E^{\gamma_{e^{+}}-\gamma_{e}} \cdot \exp\left({-E \over E_{cut_d}}\right)\right) \label{posflx} \eeq  and

The independent parameters $BF$ ,  $C_{e}$ , $\gamma_{e}$ , ${C_{s} \over C_{e}}$ , ${\gamma_{s}-\gamma_{e}}$ , $E_{cut_s}$, ${C_{e^{+}} \over C_{e}}$ , ${\gamma_{e^{+}}-\gamma_{e}}$, $E_{cut_d}$ define this model, together with $\Phi_{DM}(E)$  which depends on the Dark Matter candidate and is calculated using DarkSUSY \cite{Gondolo:2004sc} assuming NFW profile \cite{Navarro:1996gj} and a local Dark Matter density $\rho_0 = 0.3$ GeV cm$^{-3}$. As the signal mostly orginates from a region of a few kpc around the Solar system, the choice of halo model is not crucial, and the results for signal and sensitivity scale with $\rho_0^2$. The calculated Dark Matter spectra are smeared with a Gaussian distribution of 2\% width to take the energy resolution of CALET into account.  

\begin{figure}[t]

\begin{minipage}[b]{0.49\linewidth} 
\centering    
      \resizebox{\linewidth}{!}{\includegraphics{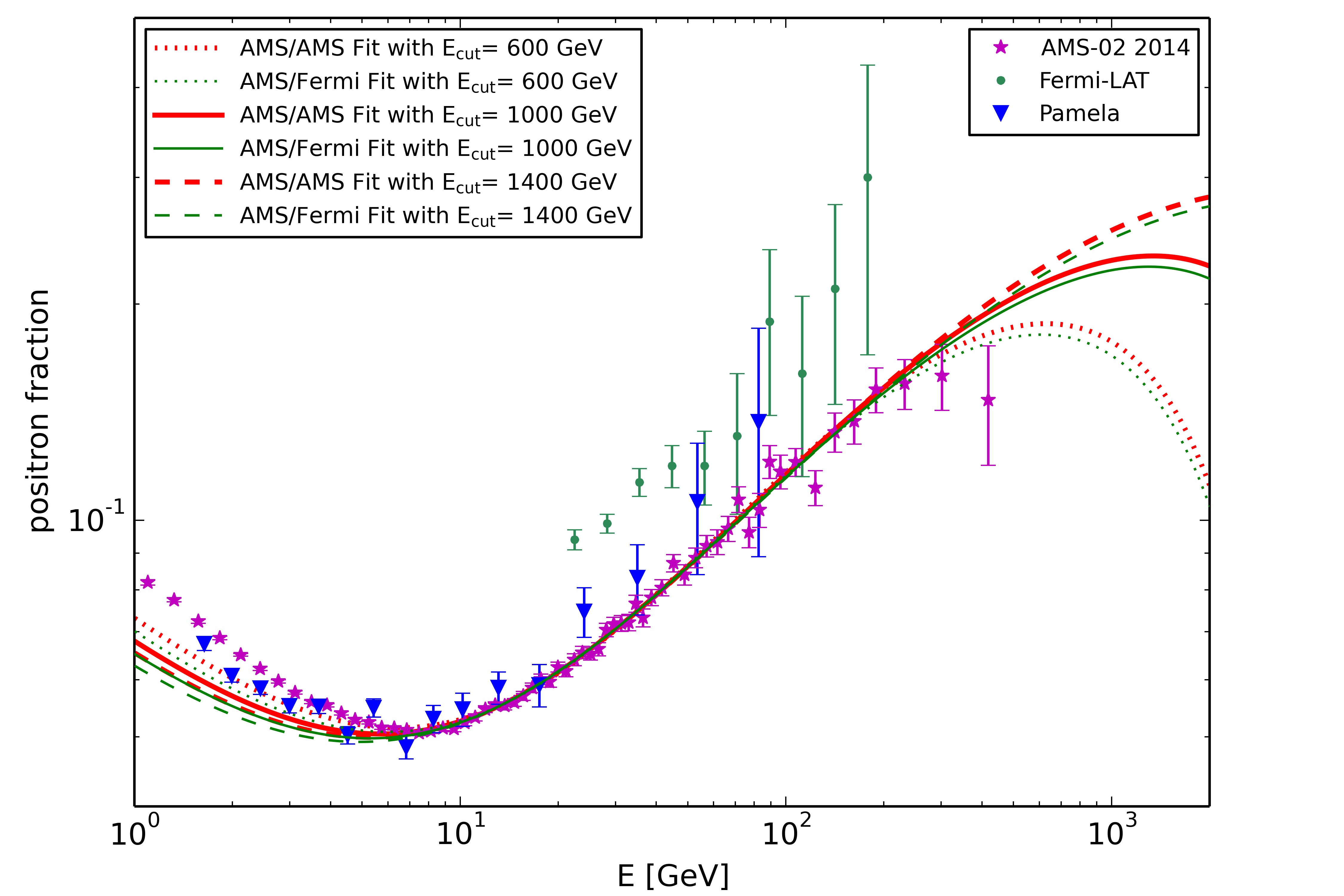}}   
\end{minipage}
\begin{minipage}[b]{0.49\linewidth} 
\centering    
      \resizebox{\linewidth}{!}{\includegraphics{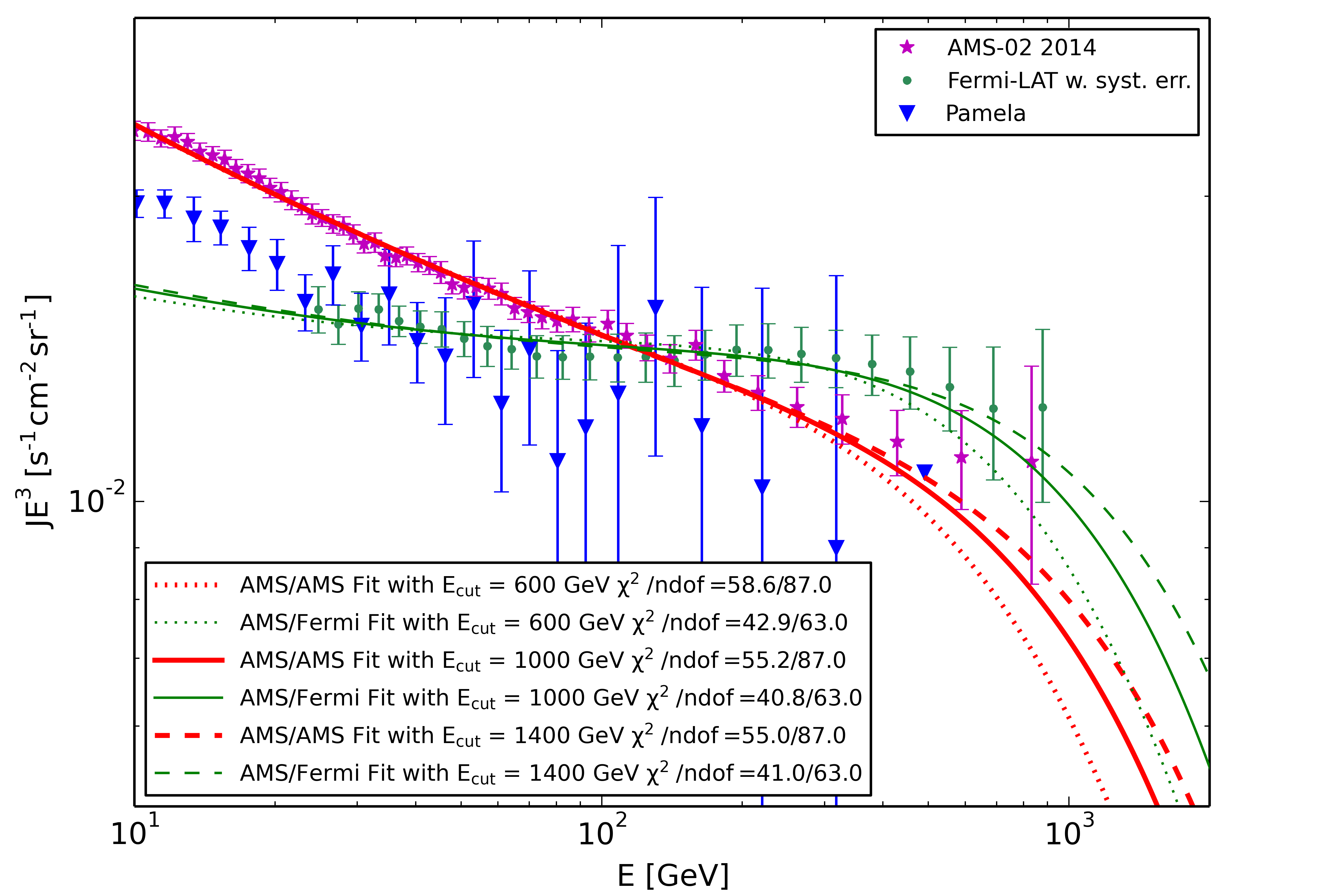}}   
\end{minipage}

\begin {center}
\caption{Best fit results for the single pulsar extra source to AMS-02 positron fraction data in combination with AMS-02 total flux data (AMS/AMS-Fit) and Fermi-LAT total flux data (AMS/Fermi-Fit). As shown in the legend, the extra source cut-off energy $E_{cut}$ does not significantly change \X2, and therefore cannot be well determined by the fit for either case. ${\gamma_{e^{+}}-\gamma_{e}}$ is set to -0.4 . \label{gssfit}} 

\end {center}
\end{figure}

In order to determine the background models for the sensitivity calculation, a combined fit of Formula (\ref{posfract}) to the AMS-02 positron fraction, and Formula (\ref{totflx}) to the AMS-02 or Fermi-LAT total flux \cite{Ackermann:2010ij} data is done. Because charge-dependent solar modulation changes the measured flux and positron fraction below 10 GeV \cite{2013PhRvL.110h1101M}, only data-points above 10 GeV were used in the fit. The cut-off energy $E_{cut_d}$ cannot be determined well by the fit, since it has influence only at high energy. It was fixed to 2 TeV and the applicability of the exponential term confirmed by numerical simulation results (described in section \ref{DRAGONsim}). The fit quality also changes only marginally with $E_{cut_s}$. This is shown in Figure \ref{gssfit} for three discrete values (0.6 TeV, 1.0 TeV and 1.4 TeV), which are accordingly used as independent background cases in the sensitivity calculations. The parameter ${\gamma_{e^{+}}-\gamma_{e}}$, which describes the slope of the positron fraction without additional source, has only significant influence below 10 GeV, making it also unobtainable by the fit. With the secondary positrons originating from interaction of primary protons with the interstellar medium, ${\gamma_{e^{+}}-\gamma_{e}}$ is close to $-\delta$, the exponent determining the diffusion coefficient's energy dependence. By this relation, its range is constrained -0.3 to -0.7, and the discrete values -0.3 , -0.4 , -0.5 , -0.6 and -0.7 were studied independently.  

\begin{figure}[t]

\begin{minipage}[b]{0.99\linewidth} 
\centering    
      \resizebox{\linewidth}{!}{\includegraphics{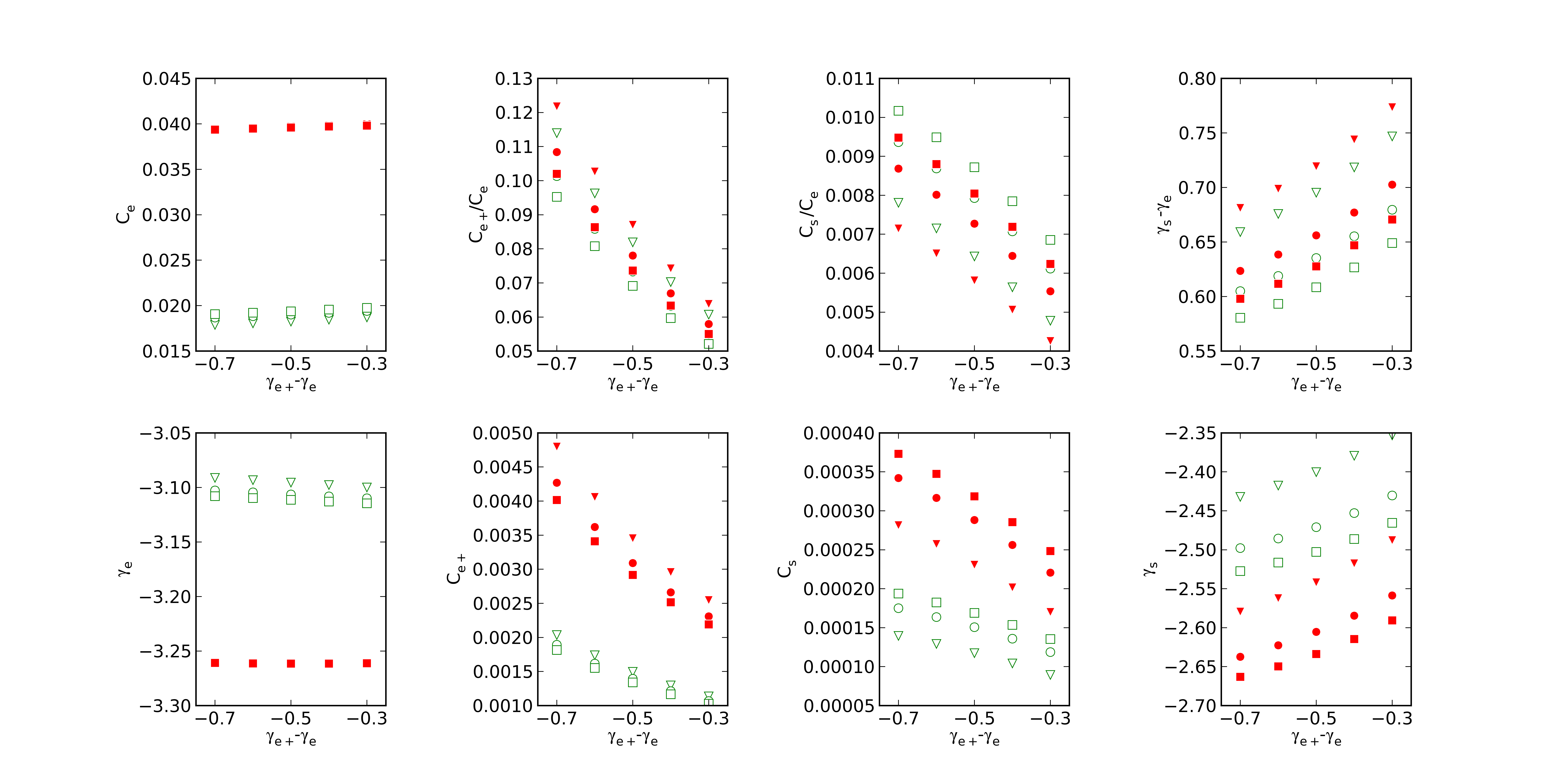}}   
\end{minipage}

\begin {center}
\caption{The fitted parameters for the single pulsar case as a function of the nuisance parameter ${\gamma_{e^{+}}-\gamma_{e}}$. Red (solid) markers represent the AMS/AMS-fit, green (hollow) markers the AMS/Fermi-fit. $E_{cut_s}$ is signified by the marker type: 0.6~TeV=$\mathrm{\bigtriangledown}$ , 1.0~TeV=$\mathrm{\Box}$ , 1.4~TeV=$\mathrm{\bigcirc}$. \label{GSSparam}} 

\end {center}
\end{figure}

The single pulsar scenario without Dark Matter gives a good fit well below 95\%CL exculsion for all of the studied fixed values of the nuisance parameters. The fitted parameters are given in Figure \ref{GSSparam}, showing that the AMS-02 total flux measurement favours a softer background spectrum index $\gamma_{e}$ with accordingly larger coefficient $C_{e}$ compared to Fermi-LAT. The positron fraction is nearly identical for both cases, as reflected by similar values for the relative fit parameters ${C_{s} \over C_{e}}$ , ${\gamma_{s}-\gamma_{e}}$ , ${C_{e^{+}} \over C_{e}}$. As a consequence, also the extra source term $\gamma_{s}$ of the AMS/AMS-fit is softer and $C_{s}$ larger than for the AMS/Fermi-fit.

\section{Method of Dark Matter Sensitivity Calculation}

As starting point for the Dark Matter sensitivity calculation serve the fits of the single pulsar case without the Dark Matter term to AMS-02 data shown in Figure \ref{gssfit}. To these background cases, the Dark Matter term is added both in total flux and positron fraction, and $BF$ increased in steps until the resultant \X2 is larger than the critical 95\%CL \X2 value for the number of free parameters in the fit. After each increase of $BF$, all free parameters of the parametrization are optimized again to adapt them to the newly added Dark Matter component. Contrary to the fit of the background case, $E_{cut_s}$ is also treated as a free parameter. To obtain the required precision at acceptable calculation time, $BF$ is increased in steps of 500 at first. Once surpassing the 95\%CL threshold, $BF$ and the fit paramters are reset to the values of the previous fit, from where $BF$ is increased in steps of 25. In the same way the step size of $BF$ is decreased to 1 and finally 0.05 with the boost factor of the final fit with \X2 below 95\%CL being the predicted limit result.

The large number of free parameters causes several pitfalls which had to be considered in devising the fitting procedure. The best value of \X2 may not depend significantly on one or more of the parameters, but not a true and unique \X2 minimum, but any set of parameters with \X2 below 95$\%$CL confirms that the assumed value of $BF$ is not excluded. Since the "Migrad" algorithm of Minuit may fail if if there is no true minimum in all dimensions, the fitting resorts to the "Simplex" algorithm in that case, avoids aborting the fit due to non-unique minima. As both algorithms follow largely the gradient of the minimized function, attention has to be paid to the fit's starting point.
In each step, the solution of the last fit with slightly smaller Dark Matter contribution is used as one start point, but another fit starting from the initial best fit parameters without Dark Matter is attempted, chosing the one with the lowest \X2. By this, the stable solution of an steady increase of the Dark Matter component is followed, while allowing the parameters to jump to a different solution if it provides a better fit. To ward against reporting too stringent limits due to a failed fit, it is confirmed that \X2 of the final fit is within 5\% the 95\percent CL threshold, which indicates a steady increse with $BF$. The limits on $BF$ obtained by this method are translated into limits on speed-averaged x-section $<\sigma v>$ by multiplication with $<\sigma v> = 3\cdot10^{-26} \mathrm{cm^3 s^{-1}}$, the value for which the Dark Matter flux $\Phi_{DM}$ is calculated. 

To simulate CALET data, statistical fluctuations to the expected event rates were taken into account by generating 100 event samples for the assumed single pulsar case. Event energies are randomly generated with the distribution of the predicted signal for 5 years of data-taking, which assumes an aperture of 1200 $\mathrm{cm^2\cdot sr}$ for CALET \cite{akaike-ICRC} and a reconstruction efficiency of 90\%. This is done by using a very fine binning of 1000 bins per decade for which the number of events in each bin is randomly selected using a Poisson distribution around the expected number of events $n$ (Gaussian if $n$ > 50), and each event assigned a random energy within the bin's boundaries. The energies of the events in each sample are then filled into energy bins again to create a dataset equivalent to what will be expected from analysis of the actual CALET data. For the data samples used in this study, the binning is 20 bins per decade for a total of 60 bins from 10 GeV to 10 TeV, which was estimated to be suitable for the applied \X2 analysis.

The expected limits are determined by performing the above described fitting procedure for each sample, and taking the average $BF$ value. Also limits from current data are calculated using the AMS-02 total flux measurement, in order to calculate the improvement expected through the addition of CALET data, since published limits using similar methods \cite{Bergstrom:2013jra}\cite{Ibarra:2013zia} differ in various assumptions, most notably the value of $\rho_0^2$, background parametrization and used data ranges.

%fails due to the found minimum not being a true minimum in all dimensions, as is confirmed by calculating the second derivaties, the fit is repeated using the "Simplex" minimizer, which does not attempt this check. This ensures that the obtained limit is not due to no unique fit solution existing for a given set of parameters.

%

%With four times higher number of events expected to be measured by AMS-02 till 2021 compared to the results published in \cite{PhysRevLett.113.121101}, which are based on 2.5 years, the statistical error of the measurement is expected reduced by half.
%Under the assumption that the current best fit with a single power law extra source is the actual physical scenario, a predicted positron fraction dataset for AMS-02 was created. To retain the \X2 of the current single power law fit, all data-points were moved half their residual towards the fit result. 

%This dataset was used in combination with the simulated CALET data to predict the expected Dark Matter limits for the year 2021.

\section{Verification of Parametrization and Methods}

\subsection{Comparison of the Parametrization to Propagation Simulation Results}
\label{DRAGONsim}

\begin{figure}[b]

\begin{minipage}[b]{0.49\linewidth} 
\centering    
      \resizebox{\linewidth}{!}{\includegraphics{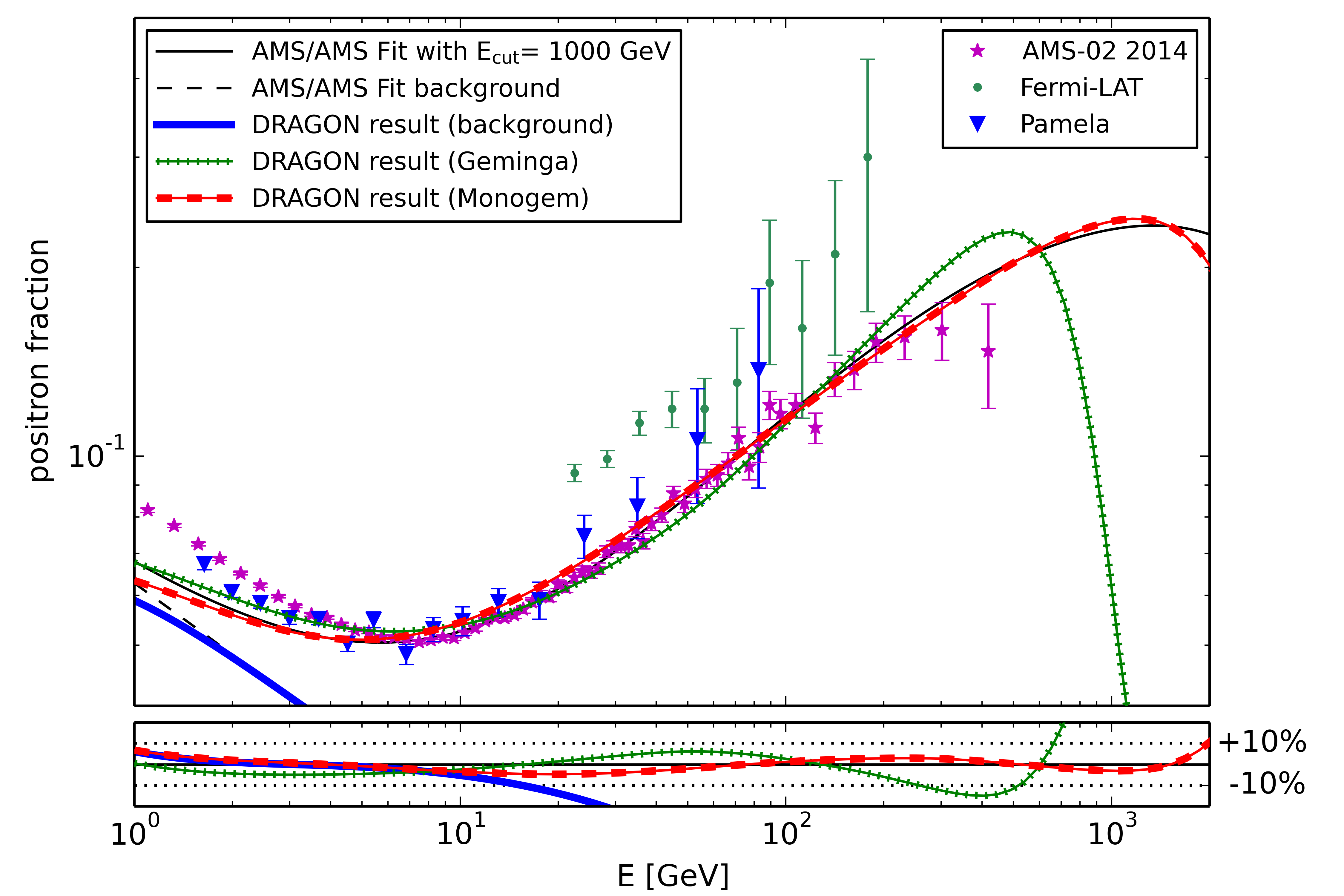}}   
\end{minipage}
\begin{minipage}[b]{0.49\linewidth} 
\centering    
      \resizebox{\linewidth}{!}{\includegraphics{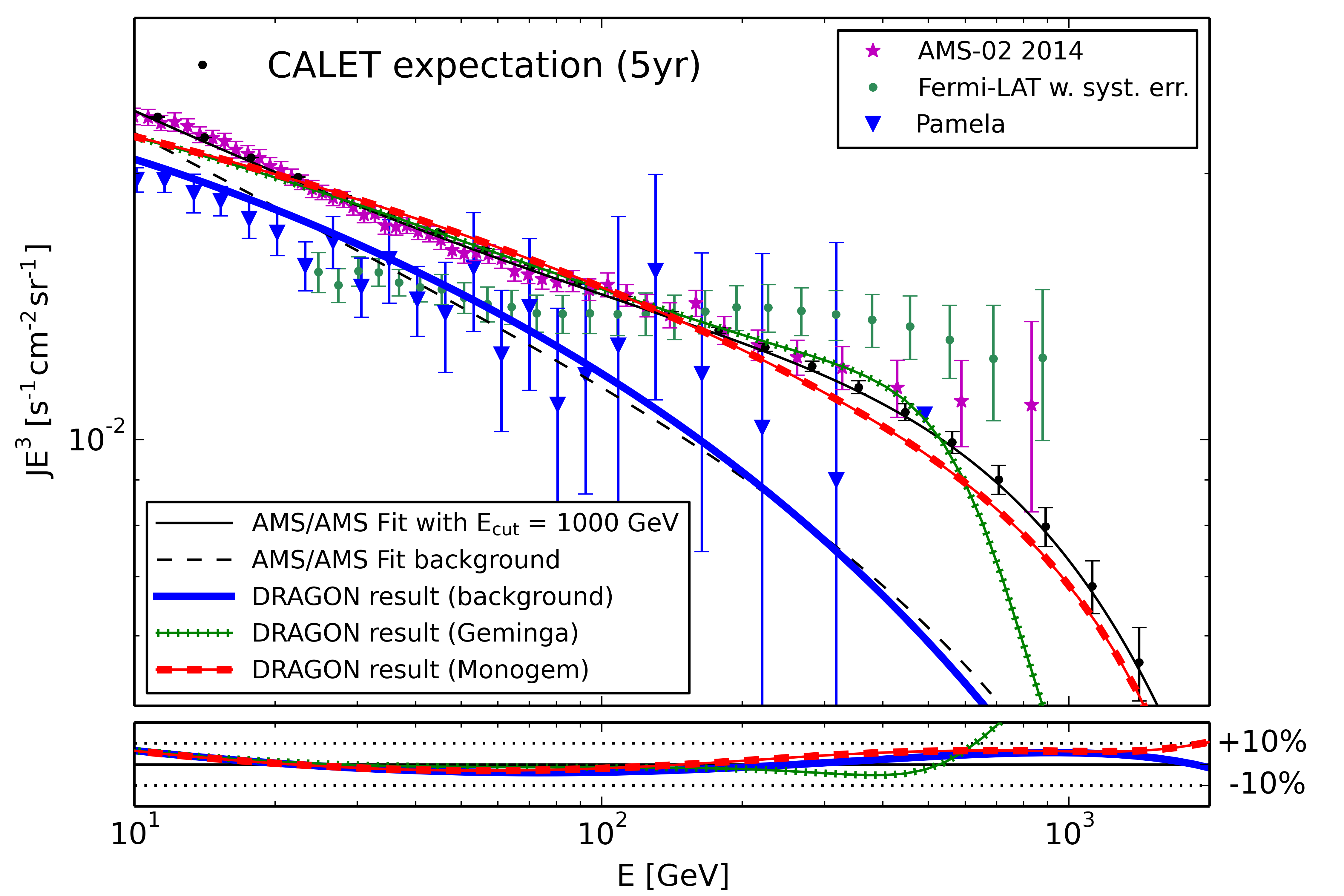}}   
\end{minipage}

\begin {center}
\caption{Best fit results for the single pulsar extra source ($E_{cut_s}$ = 1~TeV) to AMS-02 positron fraction data and AMS-02 total flux data compared to the results of the DRAGON program. In the lower panel the fractional difference $(Parametrization-DRAGON)/Parametrization$ is plotted, showing that in the most relevant energy ranges the agreement is better than 10\% . \label{dragonfit}} 

\end {center}
\end{figure}

To ensure that the adopted parametrization is in accordance with current models of cosmic ray propagation, simulations with DRAGON \cite{Gaggero:2013rya} were done. DRAGON was chosen instead of GALPROP \cite{galprop}, because it features a non-equidistant spatial grid and furthermore allowed the primary electron source distribution to be modeled according to the galaxy's spiral arm structure described in Ref.~\cite{Blasi:2011fm}. 
The injection spectrum index for electrons $\gamma_{i_e}$ is choosen equal to the index for nucleons $\gamma_{i_n}$ by default, which implies ${\gamma_{e^{+}}-\gamma_{e}}$ equal to -$\delta$. To reproduce the proton spectrum measured by PAMELA \cite{2011Sci...332...69A}, $\gamma_{i_n} = - 2.8 + \delta$ is required, where $\delta$ is the exponent in the diffusion coefficient's energy relation, $D=D_0 \cdot (E/E_0)^{\delta}$, with $E_0 = 4$ GeV. For $D_0$, values from $2.7 \cdot 10^{28}\ \mathrm{cm^2/s}$ to $7.2 \cdot 10^{28}\ \mathrm{cm^2/s}$ were considered. Under these conditions, good agreement for the background as shown in Figure \ref{dragonfit} between DRAGON and the parametrization fit to AMS-02 data equally in positron fraction and total flux is only given for ${\gamma_{e^{+}}-\gamma_{e}}$ = $-\delta$ = -0.4 ($D_0 = 6.2 \cdot 10^{28}\ \mathrm{cm^2/s}$), which was therefore chosen as the default case. The exponential cut-off due to energy loss in the parametrization is confirmed, with best agreement if the parameter $E_{cut_d}$ is set to 2~TeV. 
  
The astrophysical extra source case was implemented by simulating the Geminga pulsar wind nebula (PWN) at a distance of 0.25 kpc and with an age of 342 kyr and the Monogem PWN at a distance of 0.28 kpc and an age of 86 kpc, using the information from the ATNF catalog \cite{Manchester:2004bp}. Following \cite{Malyshev:2009tw} and \cite{2010ApJ...710..958K}, the accelerated particles are assumed to be initially trapped in a pulsar wind nebula (PWN) and accumulated over the lifetime of the PWN, for which two values (10~kyr,40~kyr) were considered. When the PWN dissolves, they are released with the intensity attenuating exponentially, the decay constant being 10~kyr.
From a scan over the injection power law index $\gamma_{i_s}$ in steps of 0.1 and cut-off energy $E_{cut_i}$ at values 1 TeV, 3 TeV and 10 TeV, best agreement is given for $\gamma_{i_s}$ = 2.1 and  $E_{cut_i} = 10$ TeV for Geminga with a lifetime of 40~kyr, and $\gamma_{i_s}$ = 2.3, $E_{cut_i}= 3$ TeV for Monogem with a PWN lifetime of 10~kyr. For these calculcations, the galaxy within in a region of 32~x~32~x~12~kpc was simulated on a three-dimensional grid with a basic grid size of 0.5~kpc. As this grid is too coarse to reliably simulate propagation in the vicinity of the solar system, a non-equidistant grid was used. For calculations of the background from distant supernovae, the grid size is reduced in steps down to 0.05 kpc near the solar system, for calculation of the nearby point source, it is reduced to 0.01 kpc near the source and the solar system.  

While the spectrum of the older Geminga pulsar shows a cut-off from radiative energy loss upward of 500~GeV, is the Monogem spectrum in full agreement with the parametrization assuming $E_{cut_s}$ = 1~TeV, demonstrating that though the spectrum in the TeV region is unknown yet, the parametrized single pulsar case is a viable scenario.

\subsection{Influence of Propagation Parameters on the Annhilation Spectra}

\begin{figure}[bh]

\begin{minipage}[b]{0.49\linewidth} 
\centering    
      \resizebox{\linewidth}{!}{\includegraphics{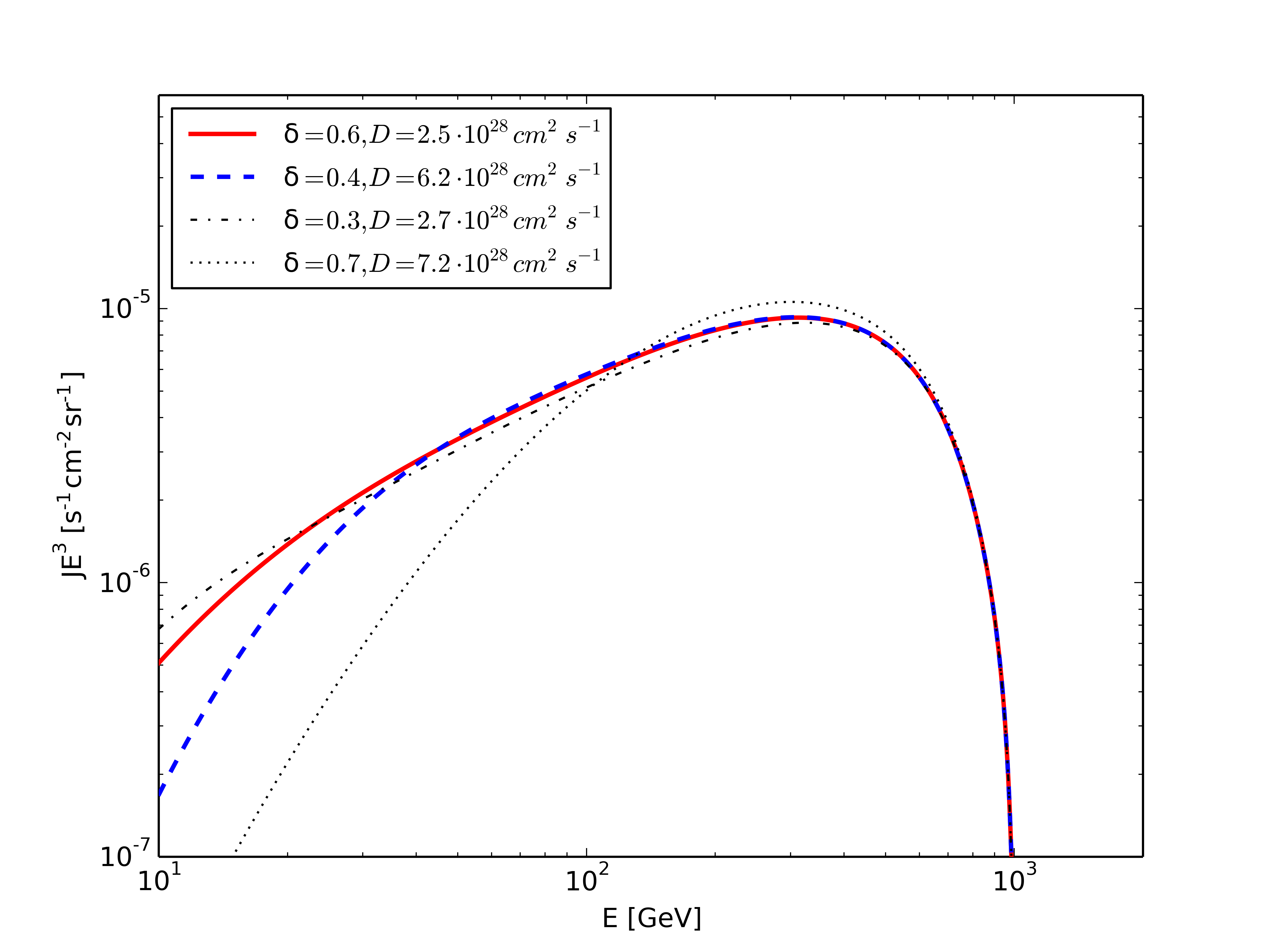}}   
\end{minipage}
\begin{minipage}[b]{0.49\linewidth} 
\centering    
      \resizebox{\linewidth}{!}{\includegraphics{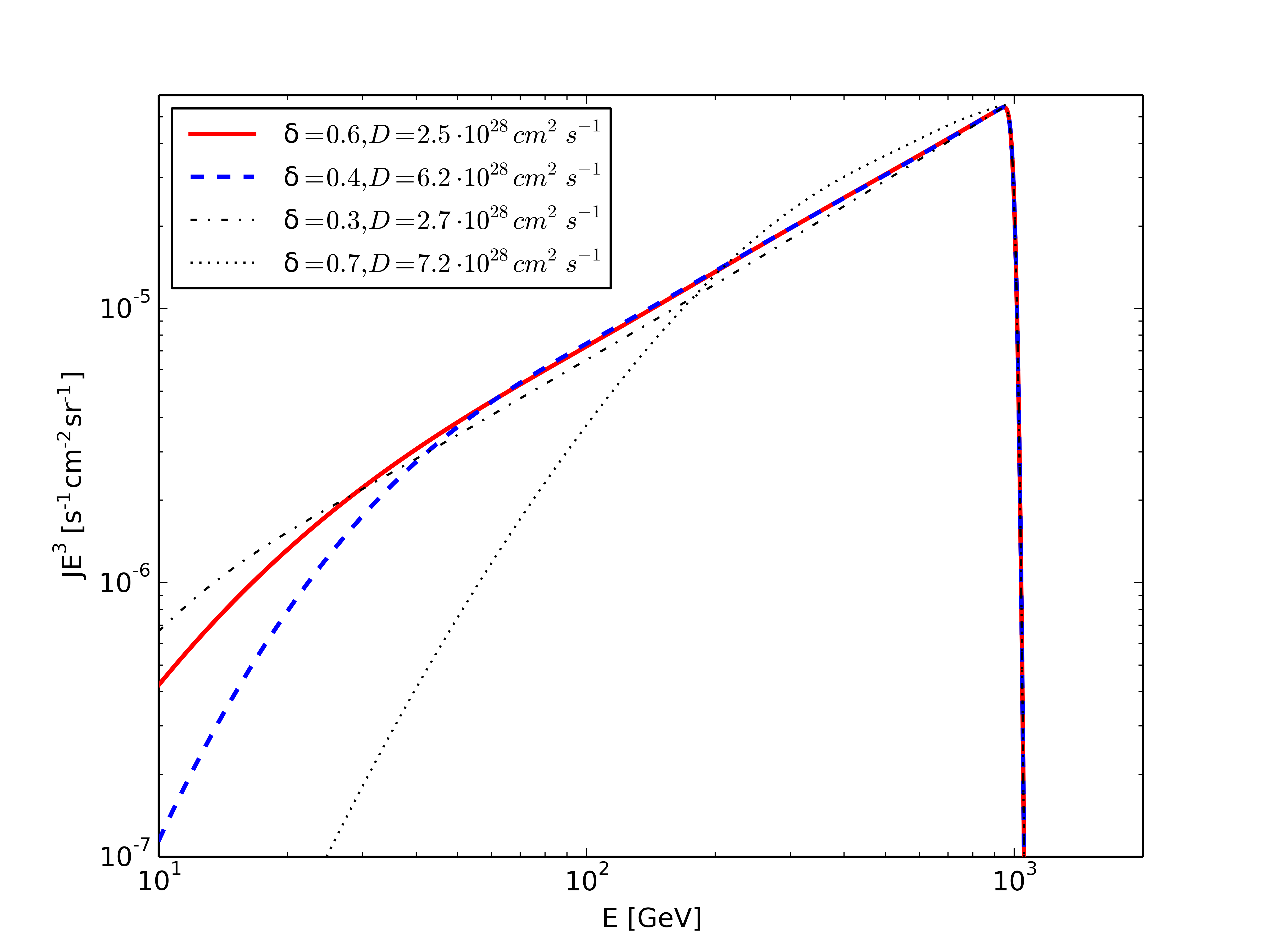}}   
\end{minipage}

\begin {center}
\caption{Comparison of the propagated Dark Matter spectra from DarkSUSY for the 100\% $\mu\bar{\mu}$-channel (left) and 100\% $e^{+}e^{-}$-channel (right). The default DarkSUSY setting (red, solid) used in the limit calculation shows above 30 GeV only marginal deviation from the case with $\delta$ = 0.4 , $D_0 = 6.2 \cdot 10^{28}\ \mathrm{cm^2/s}$ (blue, dashed), found to have the best match between the parametrization and DRAGON results. \label{DMdragon}} 

\end {center}
\end{figure}

\begin{figure}[th]

\begin{minipage}[b]{0.99\linewidth} 
\centering    
      \resizebox{\linewidth}{!}{\includegraphics{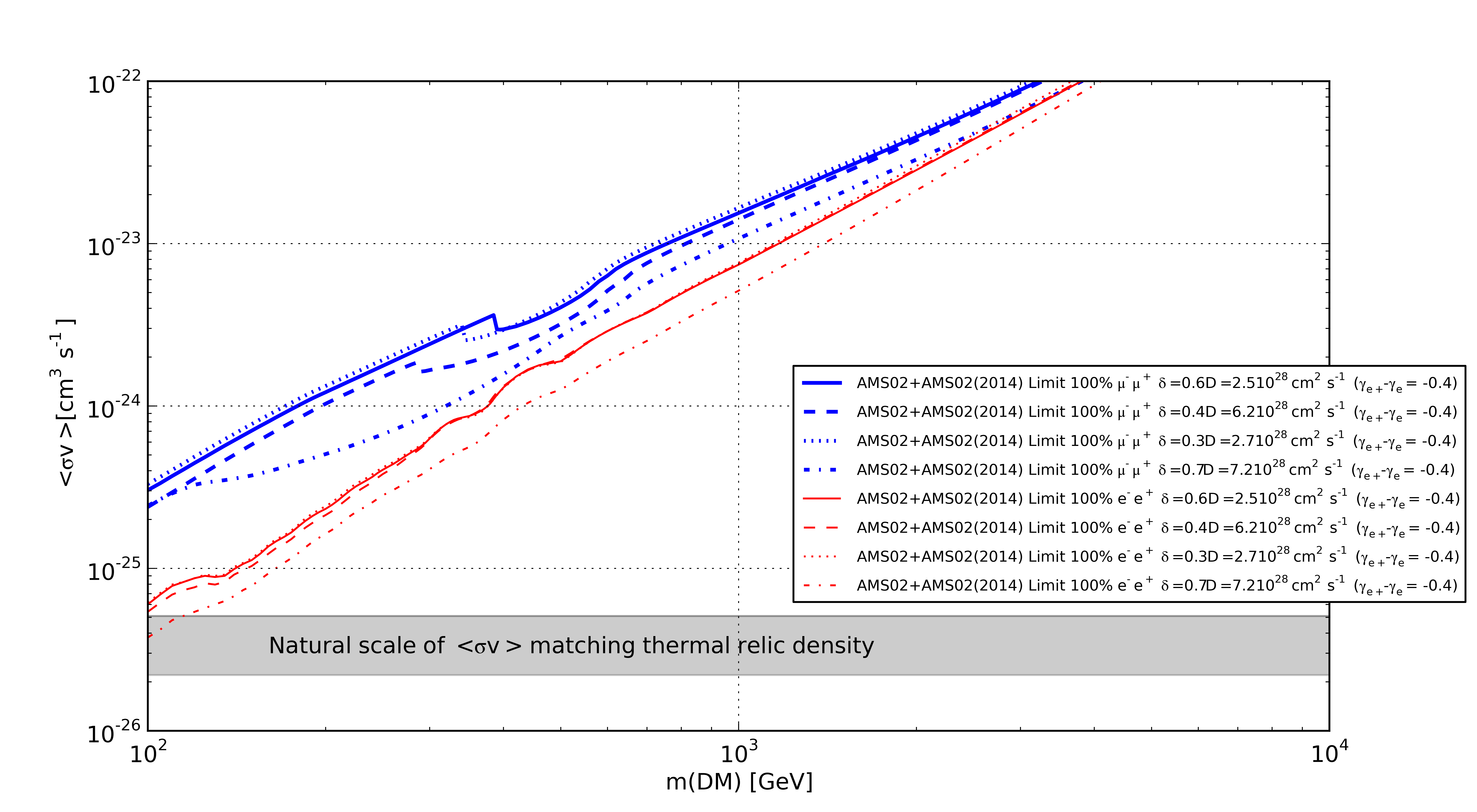}}   
\end{minipage}
\begin {center}
\caption{Comparison of the current limits from AMS02 (both positron fraction and total flux) for the 100\% $\mu\bar{\mu}$-channel (blue, thick lines) and 100\% $e^{+}e^{-}$-channel (red, thin line). The default DarkSUSY setting (solid) used in the limit calculation gives a conservative limit compared to the case with $\delta$ = 0.4 , $D_0 = 6.2 \cdot 10^{28}\ \mathrm{cm^2/s}$ (dashed) as well as within the considered range of propagation parameters, with extreme cases $\delta$ = 0.3 , $D_0 = 2.7 \cdot 10^{28}\ \mathrm{cm^2/s}$ (dotted) and $\delta$ = 0.7 , $D_0 = 7.2 \cdot 10^{28}\ \mathrm{cm^2/s}$ (dash-dot) shown. \label{proplim}} 

\end {center}
\end{figure}

The Dark Matter flux is taken from DarkSUSY, which in the used default setting treats particle propagation analytically with $\delta$ = 0.6 and $D_0$ = 2.5 $\cdot 10^{28} cm^2/s$.  It is shown in Figure \ref{DMdragon}, that variation of the propagation parameters has little influence on the propagated annihilation spectra except for low energy.
As demonstrated by Figure \ref{proplim} on the example of the current AMS-02 limits, the used default setting gives a rather conservative limit, very close to that of the $\delta$ = 0.4 , $D_0 = 6.2 \cdot 10^{28}\ \mathrm{cm^2/s}$ case.

%The agreement of these spectra with results from DRAGON was tested, also investigating the influence of the propagation parameters.

\section{Expected Limits on Dark Matter Annihilation from CALET Data}

\begin{figure}[b]

\begin{minipage}[b]{0.99\linewidth} 
\centering    
      \resizebox{\linewidth}{!}{\includegraphics{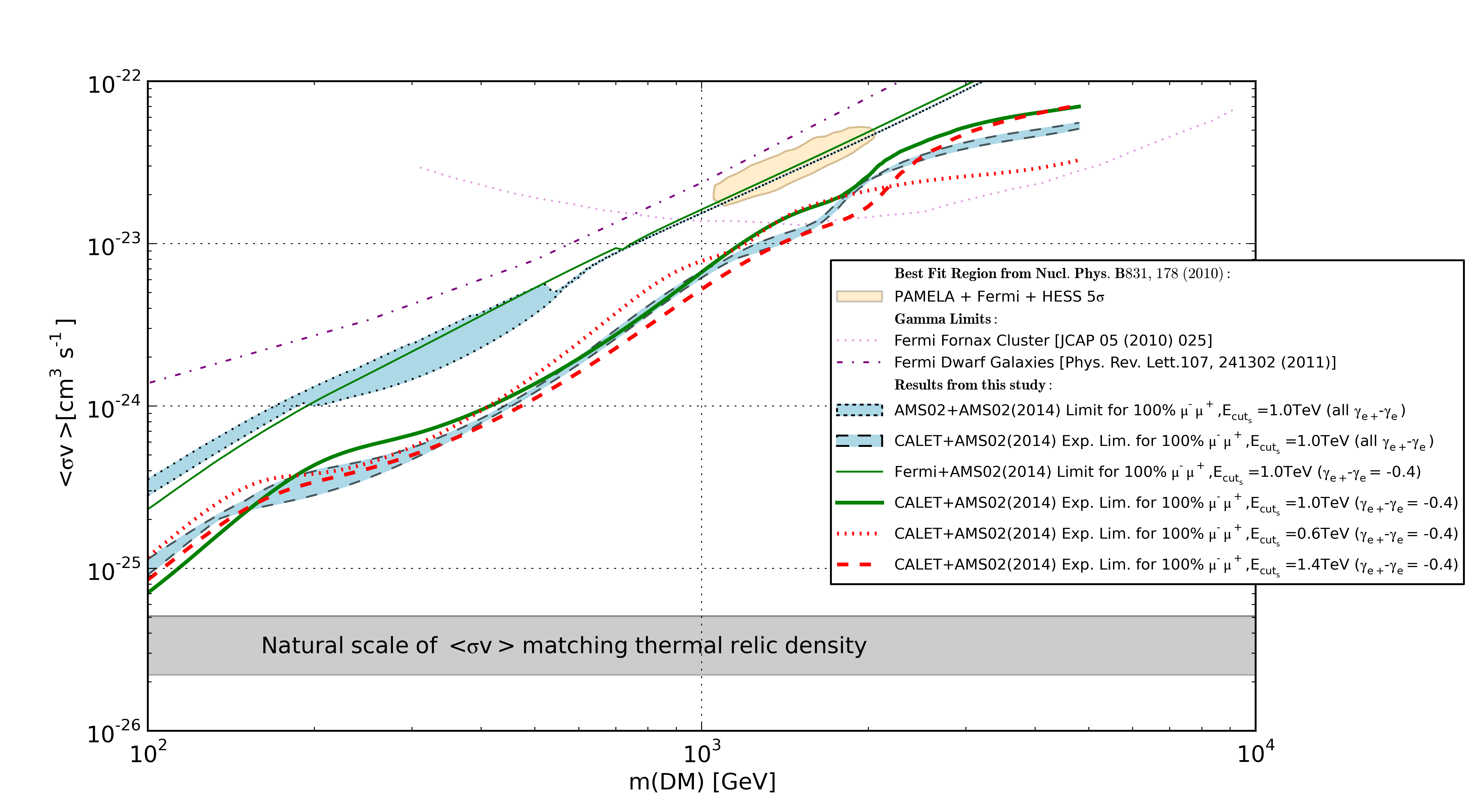}}   
\end{minipage}

\begin {center}
\caption{Expected Dark Matter annihilation limits (100\% $\mu\bar{\mu}$-channel) from CALET for different background cases. The blue area with dashed (dotted) border shows the range for varying ${\gamma_{e^{+}}-\gamma_{e}}$ from -0.3 to -0.7 in the AMS/AMS-Fit background case for expected (current) limits. The dotted (dashed) red line shows the limit for a source cut-off energy $E_{cut_s}$ of 0.6~TeV (1.4~TeV), and the solid green thick (thin) line represents the expected (current) limit for the AMS/Fermi background case. As a reference point, the best fit region from Ref. \cite{Meade:2009iu} for the positron excess being caused only by Dark Matter annihilation is shown, as well as previous limits on the $\mu\bar{\mu}$-channel from $gamma$-observations with Fermi \cite{2010JCAP...05..025A} \cite{PhysRevLett.107.241302}.\label{FPSdep}} 

\end {center}
\end{figure}

\begin{figure}

\begin{minipage}[b]{0.99\linewidth} 
\centering    
      \resizebox{\linewidth}{!}{\includegraphics{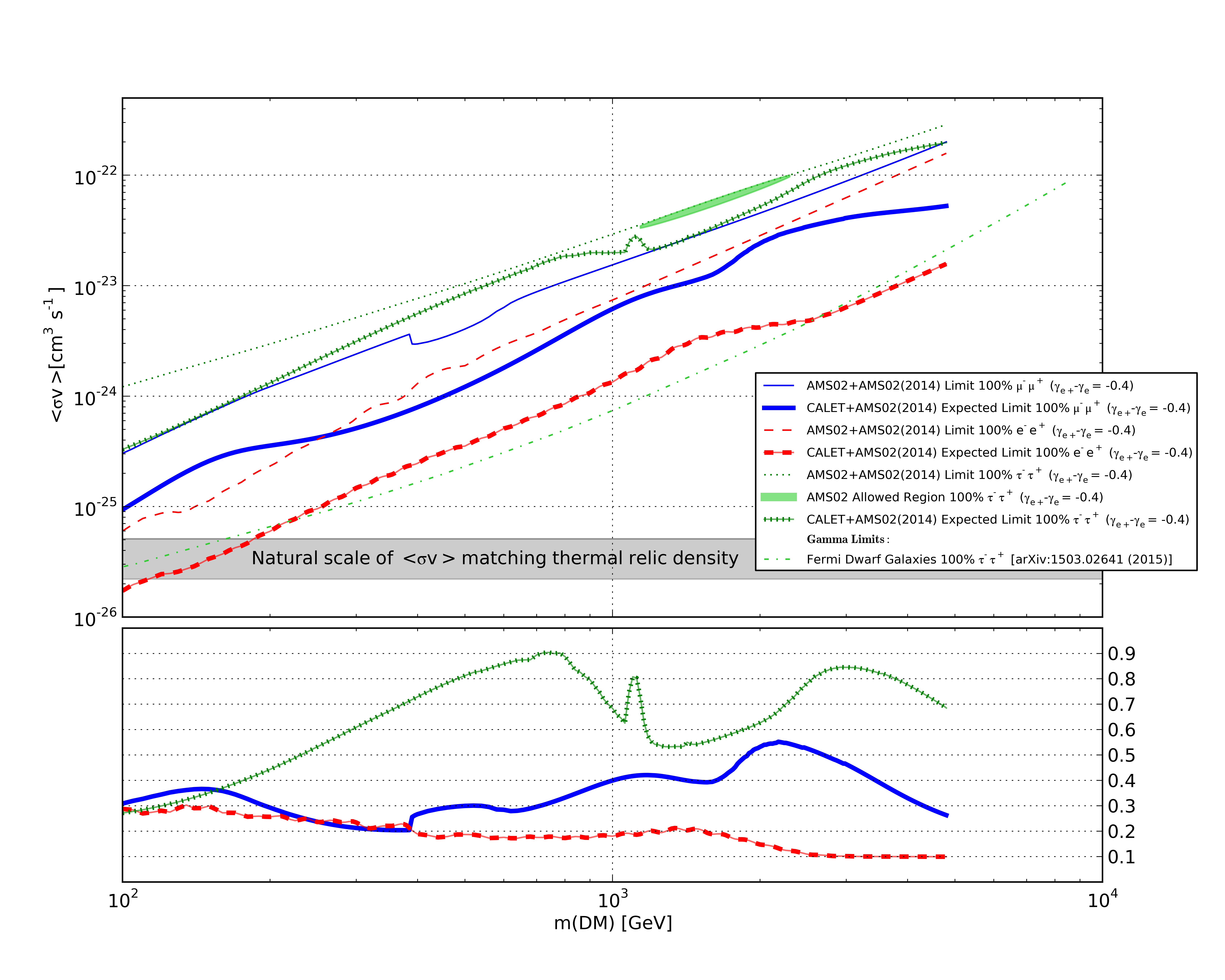}}    
\end{minipage}

\begin {center}
\caption{Current (AMS-02) and expected (AMS-02 + CALET) annihilation x-section limits for pure annihilation into the lepton-channels assuming $E_{cut_s} = 1$ TeV. For $\tau^+\tau^-$ -annihilating Dark Matter, current AMS-02 data shows an "Allowed Region" where the excess is caused only by Dark Matter annihilation. The lower panel shows the ratio of the current and the predicted limits.\label{leptchan} } 

\end {center}
\end{figure}

In Figure \ref{FPSdep}, the current AMS-02 limit and predicted AMS-02 + CALET limits under different assumptions ar shown for the 100\% $\mu\bar{\mu}$-channel, which was selected as a standard case due to its averagely hard/soft spectrum and the availability of reference results from other publications \cite{Meade:2009iu}\cite{2010JCAP...05..025A}\cite{PhysRevLett.107.241302}. The expected limits have no significant dependence on ${\gamma_{e^{+}}-\gamma_{e}}$. Their shape depends on the initial value of $E_{cut_s}$ as it influences for which mass the Dark Matter spectrum is closest to the assumed pulsar spectrum. Choosing the AMS/Fermi background case instead of AMS/AMS results in a slightly different, but comparable predicted limit. 

%All these limits would surpass those of indirect Dark Matter search in the $\gamma$-ray spectra of extragalactic sources and exclude muon-channel annihilation of Dark Matter as the origin of the positron excess.  

Figure \ref{leptchan} gives an overview of the expected limits for Dark Matter annihilation purely into leptons. It is shown that by combination of CALET data with AMS-02 positron fraction, the limits on Dark Matter annihilation can be improved for all studied Dark Matter candidates, which is foremost attributed to better statistics of the CALET total flux data over the whole energy range. 
%The limits derived from the AMS-02 positron flux in \cite{Ibarra:2013zia} are shown for comparison, rescaled by the squared ratio of the assumed local Dark Matter densities, which is $0.39^2 / 0.3^2 = 1.69$. These limits take the optimum limit from a sample of energy windows, while we use the whole energy range available. Taking the variation of the energy window into account, our limits are comparable. It should also be noted that the used parametrizations are fundamentally different in that we assume an extra source emitting an equal amount of electrons and positrons, and also achieve an reasonable prediction for the TeV region flux matching numerical proagation calculation results. 
The similiarity of the $\tau^+\tau^-$-channel's spectrum to the pulsar spectrum results in a generally weak limit with a dent just above 1 TeV, where the combination of Dark Matter annihilation and a modified pulsar spectrum matches the assumed spectrum best. While current AMS-02 data allows the positron excess to be fully attributed to $\tau^+\tau^-$-channel Dark Matter annihilation, the very strong limits from $\gamma$-ray observation of dwarf galaxies by Fermi-LAT \cite{Ackermann:2015zua}, despite systematic uncertainties, most likely rule out that scenario already.

\begin{figure}

\begin{minipage}[b]{0.99\linewidth} 
\centering    
      \resizebox{\linewidth}{!}{\includegraphics{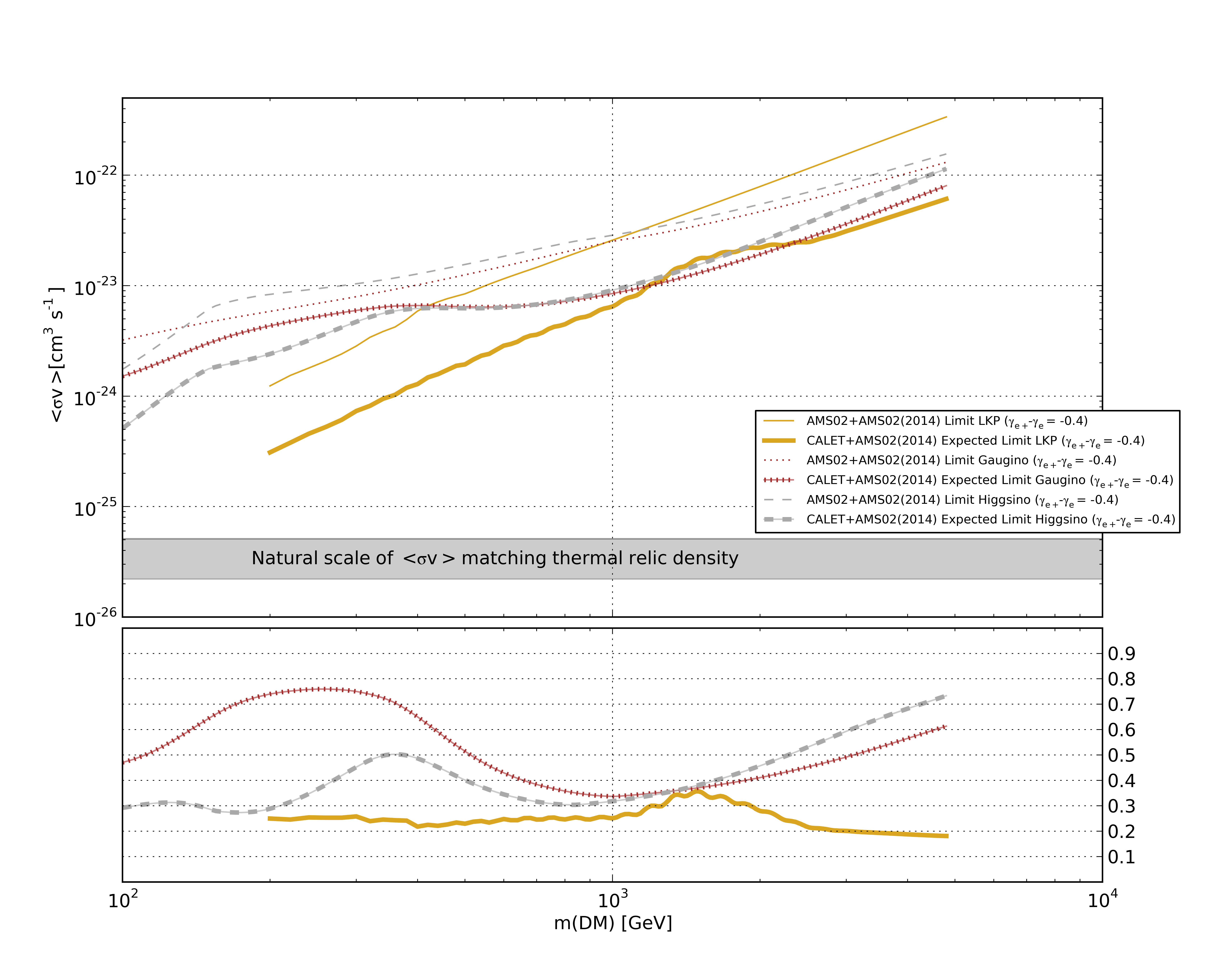}}    
\end{minipage}

\begin {center}
\caption{Current (AMS-02) and expected (AMS-02 + CALET) annihilation x-section limits for different Dark Matter candidates, assuming $E_{cut_s} = 1$ TeV. The lower panel shows the ratio of the current and the predicted limits. As the annihilation of LKP includes the $t\bar{t}$-channel, no limits below 200~GeV are calculated. \label{DMcand} } 

\end {center}
\end{figure}

Figure \ref{DMcand} shows the current and expected limits for the LKP (Lightest Kaluza-Klein Particle) Dark Matter candidate, as well as the two extreme cases from minimal supersymmetry, Gaugino and Higgsino.\\\begin{minipage}{\linewidth} \vspace{2mm} The annihilation branching ratios for these Dark Matter candidates are assumed as follows: \\ 
\begin{tabular}{llllll}
{\rule{0pt}{3ex}LKP:} & {20\% $e^+ e^-$} & {(1.2\% $\nu_e \bar{\nu_e}$)} & {(11\% $u\bar{u}$)} & {(0.7\% $d\bar{d}$)} & {(2.3\% $h\bar{h}$)} \\
{ }     & {20\% $\mu^+\mu^-$} & {(1.2\% $\nu_\mu \bar{\nu_\mu}$)}   & {11\% $c\bar{c}$} & {(0.7\% $s\bar{s}$)} & {} \\
{ }     & {20\% $\tau^+\tau^-$} & {(1.2\% $\nu_\tau \bar{\nu_\tau}$)} & {11\% $t\bar{t}$} & {0.7\% $b\bar{b}$} & {} \\

{\rule{0pt}{3ex}Higgsino:} & {50\% $W^+ W^-$} & {50\% $Z\bar{Z}$}&{}&{}\\
{\rule{0pt}{3ex}Gaugino:} & {15\% $\tau^+\tau^-$} & {85\% $b\bar{b}$} &{}&{}\\
\end{tabular}\\ \vspace{1mm}

The branching ratio for LKP was taken from \cite{Hooper:2002gs}, with the channels in brackets not contributing to the signal due to being considered marginal and thus having not been implemented in the used DarkSUSY routines for positron flux calculation \cite{Gondolo:2004sc}.
\end{minipage} \vspace{1mm}

For LKP and annihilation to $e^+ e^-$ the improvement from current limits is around half an order of magnitude for most of the studied Dark Matter mass range, for the latter reaching up to a full order of magnitude above 2 TeV. This is due to these cases featuring a sudden drop of the spectrum at the energy matching the mass of the Dark Matter particles, which causes a mismatch with the single pulsar assumption, since in contrast to AMS-02 data, the CALET data extends up to this energy with good statistics and energy resolution. 
The Supersymmetric Dark Matter candidates exhibit also a strong expected improvement of the limit for a Dark Matter mass above 1~TeV, at which the high number of comparatively low energy electrons from hadronic showers in the decay of the primary annihilation products (quarks and gauge bosons) starts to create an significant excess above 10~GeV, the lower boundary of the used energy range.

\section{Conclusion}

CALET will measure the total electron+positron flux up to 20 TeV with good statistics and energy resolution even in the TeV region. Assuming the positron excess is caused by a single nearby pulsar, the limits obtainable from CALET's measurement on Dark Matter annihilation purely into leptons, as well as for LKP, Gaugino and Higgsino Dark Matter with mass from 100~GeV to 4.8~TeV have been calculated. Their reliability with regard to nuisance parameters, and the agreement of the used parametrization with numerical propagation calculation were confirmed. If CALET data is found to match the single pulsar scenario, it will be possible to set significantly more stringent limits on Dark Matter annihilation compared to current experimental data, especially for annihilation to $e^+ e^-$ and LKP.
Though strong limits from $\gamma$-ray observation of the galactic center and dwarf galaxies exist, they strongly depend on the assumption for the Dark Matter halo profile. Observation of electron and positrons in the upper GeV and TeV range is on the other hand sensitive to the local and nearby Dark Matter density, providing an important complementary way of indirect Dark Matter search.   
CALET is optimized to detect nearby supernova remnants \cite{Torii201155}, making it possible to include them in the background model for Dark Matter search when CALET data becomes available. While the presented sensitivity assumes the positron excess is not caused by Dark Matter at all, the actual CALET data may hint at Dark Matter partially contributing to the positron excess or even causing it completely, for example in the form of decaying Dark Matter \cite{Ibe:2014qya}, in which case CALET data will be analyzed for discovering the signatures of these scenarios.

%\appendix
%\section{Some title}
%Please always give a title also for appendices.

%\acknowledgments

%This is the most common positions for acknowledgments. A macro is
%available to maintain the same layout and spelling of the heading.

%\paragraph{Note added.} This is also a good position for notes added
%after the paper has been written.


\bibliographystyle{JHEP}
\bibliography{ref}

\providecommand{\href}[2]{#2}\begingroup\raggedright\begin{thebibliography}{10}

\bibitem{Torii-ICRC}
{\bf CALET} Collaboration, S.~Torii, {\it {The CALorimetric Electron Telescope
  (CALET): a High-Energy Astroparticle Physics Observatory on the International
  Space Station}},  in {\em Proceedings of the 34th International Cosmic Ray
  Conference (ICRC 2015)}, 2015.
\newblock ID 581.

\bibitem{0004-637X-772-2-108}
L.~Zhang, H.-W. Rix, G.~van~de Ven, J.~Bovy, C.~Liu, and G.~Zhao, {\it The
  gravitational potential near the sun from segue k-dwarf kinematics},  {\em
  The Astrophysical Journal} {\bf 772} (2013), no.~2 108,
  [\href{http://arxiv.org/abs/1209.0256}{{\tt arXiv:1209.0256}}].

\bibitem{2015arXiv150708581S}
H.~{Silverwood}, S.~{Sivertsson}, P.~{Steger}, J.~I. {Read}, and G.~{Bertone},
  {\it {A non-parametric method for measuring the local dark matter density}},
  {\em ArXiv e-prints} (July, 2015)
  [\href{http://arxiv.org/abs/1507.08581}{{\tt arXiv:1507.08581}}].

\bibitem{Adriani:2008zr}
{\bf PAMELA Collaboration} Collaboration, O.~Adriani et~al., {\it {An anomalous
  positron abundance in cosmic rays with energies 1.5-100 GeV}},  {\em Nature}
  {\bf 458} (2009) 607--609, [\href{http://arxiv.org/abs/0810.4995}{{\tt
  arXiv:0810.4995}}].

\bibitem{PhysRevLett.113.121101}
{\bf AMS} Collaboration, L.~Accardo et~al., {\it High statistics measurement of
  the positron fraction in primary cosmic rays of 0.5\char21{}500~gev with the
  alpha magnetic spectrometer on the international space station},  {\em Phys.
  Rev. Lett.} {\bf 113} (Sep, 2014) 121101.

\bibitem{PhysRevLett.110.141102}
{\bf AMS} Collaboration, M.~Aguilar et~al., {\it First result from the alpha
  magnetic spectrometer on the international space station: Precision
  measurement of the positron fraction in primary cosmic rays of 0.5–350
  gev},  {\em Phys. Rev. Lett.} {\bf 110} (Apr, 2013) 141102.

\bibitem{PhysRevLett.113.221102}
{\bf AMS} Collaboration, M.~Aguilar et~al., {\it Precision measurement of the
  $({e}^{+}+{e}^{-})$ flux in primary cosmic rays from 0.5~gev to 1~tev with
  the alpha magnetic spectrometer on the international space station},  {\em
  Phys. Rev. Lett.} {\bf 113} (Nov, 2014) 221102.

\bibitem{Gondolo:2004sc}
P.~Gondolo, J.~Edsjo, P.~Ullio, L.~Bergstrom, M.~Schelke, et~al., {\it
  {DarkSUSY: Computing supersymmetric dark matter properties numerically}},
  {\em JCAP} {\bf 0407} (2004) 008,
  [\href{http://arxiv.org/abs/astro-ph/0406204}{{\tt astro-ph/0406204}}].

\bibitem{Navarro:1996gj}
J.~F. Navarro, C.~S. Frenk, and S.~D. White, {\it {A Universal density profile
  from hierarchical clustering}},  {\em Astrophys.J.} {\bf 490} (1997)
  493--508, [\href{http://arxiv.org/abs/astro-ph/9611107}{{\tt
  astro-ph/9611107}}].

\bibitem{Ackermann:2010ij}
{\bf Fermi LAT} Collaboration, M.~Ackermann et~al., {\it {Fermi LAT
  observations of cosmic-ray electrons from 7 GeV to 1 TeV}},  {\em Phys.Rev.}
  {\bf D82} (2010) 092004, [\href{http://arxiv.org/abs/1008.3999}{{\tt
  arXiv:1008.3999}}].

\bibitem{2013PhRvL.110h1101M}
L.~{Maccione}, {\it {Low Energy Cosmic Ray Positron Fraction Explained by
  Charge-Sign Dependent Solar Modulation}},  {\em Phys. Rev. Lett.} {\bf 110}
  (Feb., 2013) 081101, [\href{http://arxiv.org/abs/1211.6905}{{\tt
  arXiv:1211.6905}}].

\bibitem{akaike-ICRC}
Y.~Akaike, K.~Kasahara, and S.~Torii, {\it {Expected CALET Telescope
  Performance from Monte Carlo Simulations}},  in {\em Proceedings of the 32nd
  International Cosmic Ray Conference (ICRC 2011)}, vol.~6, p.~371, 2011.

\bibitem{Bergstrom:2013jra}
L.~Bergstrom, T.~Bringmann, I.~Cholis, D.~Hooper, and C.~Weniger, {\it {New
  limits on dark matter annihilation from AMS cosmic ray positron data}},  {\em
  Phys. Rev. Lett.} {\bf 111} (2013) 171101,
  [\href{http://arxiv.org/abs/1306.3983}{{\tt arXiv:1306.3983}}].

\bibitem{Ibarra:2013zia}
A.~Ibarra, A.~S. Lamperstorfer, and J.~Silk, {\it {Dark matter annihilations
  and decays after the AMS-02 positron measurements}},  {\em Phys.Rev.} {\bf
  D89} (2014) 063539, [\href{http://arxiv.org/abs/1309.2570}{{\tt
  arXiv:1309.2570}}].

\bibitem{Gaggero:2013rya}
D.~Gaggero, L.~Maccione, G.~Di~Bernardo, C.~Evoli, and D.~Grasso, {\it
  {Three-Dimensional Model of Cosmic-Ray Lepton Propagation Reproduces Data
  from the Alpha Magnetic Spectrometer on the International Space Station}},
  {\em Phys.Rev.Lett.} {\bf 111} (2013), no.~2 021102,
  [\href{http://arxiv.org/abs/1304.6718}{{\tt arXiv:1304.6718}}]. DRAGON
  website: http://www.dragonproject.org.

\bibitem{galprop}
GALPROP website: http://galprop.stanford.edu/.

\bibitem{Blasi:2011fm}
P.~Blasi and E.~Amato, {\it {Diffusive propagation of cosmic rays from
  supernova remnants in the Galaxy. II: anisotropy}},  {\em JCAP} {\bf 1201}
  (2012) 011, [\href{http://arxiv.org/abs/1105.4529}{{\tt arXiv:1105.4529}}].

\bibitem{2011Sci...332...69A}
{\bf PAMELA} Collaboration, O.~{Adriani} et~al., {\it {PAMELA Measurements of
  Cosmic-Ray Proton and Helium Spectra}},  {\em Science} {\bf 332} (Apr., 2011)
  69--, [\href{http://arxiv.org/abs/1103.4055}{{\tt arXiv:1103.4055}}].

\bibitem{Manchester:2004bp}
R.~N. Manchester, G.~B. Hobbs, A.~Teoh, and M.~Hobbs, {\it {The Australia
  Telescope National Facility pulsar catalogue}},  {\em Astron.J.} {\bf 129}
  (2005) 1993, [\href{http://arxiv.org/abs/astro-ph/0412641}{{\tt
  astro-ph/0412641}}].

\bibitem{Malyshev:2009tw}
D.~Malyshev, I.~Cholis, and J.~Gelfand, {\it {Pulsars versus Dark Matter
  Interpretation of ATIC/PAMELA}},  {\em Phys.Rev.} {\bf D80} (2009) 063005,
  [\href{http://arxiv.org/abs/0903.1310}{{\tt arXiv:0903.1310}}].

\bibitem{2010ApJ...710..958K}
N.~{Kawanaka}, K.~{Ioka}, and M.~M. {Nojiri}, {\it {Is Cosmic Ray Electron
  Excess from Pulsars Spiky or Smooth?: Continuous and Multiple
  Electron/Positron Injections}},  {\em Astrophys.J.} {\bf 710} (Feb., 2010)
  958--963, [\href{http://arxiv.org/abs/0903.3782}{{\tt arXiv:0903.3782}}].

\bibitem{Meade:2009iu}
P.~Meade, M.~Papucci, A.~Strumia, and T.~Volansky, {\it {Dark Matter
  Interpretations of the e+- Excesses after FERMI}},  {\em Nucl. Phys.} {\bf
  B831} (2010) 178--203, [\href{http://arxiv.org/abs/0905.0480}{{\tt
  arXiv:0905.0480}}].

\bibitem{2010JCAP...05..025A}
{\bf The Fermi-LAT Collaboration} Collaboration, M.~Ackermann et~al., {\it
  {Constraints on dark matter annihilation in clusters of galaxies with the
  Fermi large area telescope}},  {\em JCAP} {\bf 5} (May, 2010) 25,
  [\href{http://arxiv.org/abs/1002.2239}{{\tt arXiv:1002.2239}}].

\bibitem{PhysRevLett.107.241302}
{\bf The Fermi-LAT Collaboration} Collaboration, M.~Ackermann et~al., {\it
  Constraining dark matter models from a combined analysis of milky way
  satellites with the fermi large area telescope},  {\em Phys. Rev. Lett.} {\bf
  107} (Dec, 2011) 241302, [\href{http://arxiv.org/abs/1108.3546}{{\tt
  arXiv:1108.3546}}].

\bibitem{Ackermann:2015zua}
{\bf Fermi-LAT} Collaboration, M.~Ackermann et~al., {\it {Searching for Dark
  Matter Annihilation from Milky Way Dwarf Spheroidal Galaxies with Six Years
  of Fermi-LAT Data}},  \href{http://arxiv.org/abs/1503.02641}{{\tt
  arXiv:1503.02641}}.

\bibitem{Hooper:2002gs}
D.~Hooper and G.~D. Kribs, {\it {Probing Kaluza-Klein dark matter with neutrino
  telescopes}},  {\em Phys. Rev.} {\bf D67} (2003) 055003,
  [\href{http://arxiv.org/abs/hep-ph/0208261}{{\tt hep-ph/0208261}}].

\bibitem{Torii201155}
S.~Torii, {\it Calorimetric electron telescope mission: Search for dark matter
  and nearby sources},  {\em NIM A} {\bf 630} (2011), no.~1 55 -- 57.
  Proceedings of RICAP 2009.

\bibitem{Ibe:2014qya}
M.~Ibe, S.~Matsumoto, S.~Shirai, and T.~T. Yanagida, {\it {Mass of Decaying
  Wino from AMS-02 2014}},  {\em Phys.Lett.} {\bf B741} (2015) 134--137,
  [\href{http://arxiv.org/abs/1409.6920}{{\tt arXiv:1409.6920}}].

\end{thebibliography}\endgroup

%\begin{thebibliography}{99}
%\bibitem{a}
%Author, \emph{Title}, \emph{J. Abbrev.} {\bf vol} (year) pg.
%\bibitem{b}
%Author, \emph{Title},
%arxiv:1234.5678.
%\bibitem{c}
%Author, \emph{Title},
%Publisher (year).
% Please avoid comments such as "For a review'', "For some examples",
% "and references therein" or move them in the text. In general,
% please leave only references in the bibliography and move all
% accessory text in footnotes.
% Also, please have only one work for each \bibitem.
%\end{thebibliography}
\end{document}